\documentstyle[prb,aps,epsf,floats]{revtex}

\begin{document}
\draft

\twocolumn[\hsize\textwidth\columnwidth\hsize\csname@twocolumnfalse%
\endcsname

\title{Domain-Wall Free-Energy of Spin Glass Models : \\
NumericalMethod and Boundary Conditions }
\author{Koji Hukushima}
\address{Institute for Solid State Physics, Univ. of Tokyo, 7-22-1
  Roppongi, Minato-ku, Tokyo 106-8666,  Japan}
\date{\today}
\maketitle
\begin{abstract}
An efficient Monte Carlo  method is extended to evaluate directly
domain-wall 
free-energy for randomly frustrated spin systems. 
Using the method, critical phenomena of spin-glass phase transition
is investigated in $4d \pm J$ Ising model under the replica boundary
condition. 
Our values of the critical temperature and exponent, obtained by
finite-size scaling, are in good agreement with those of the standard
MC and the series expansion studies. In addition, 
two exponents, the stiffness exponent and the fractal dimension of the 
domain wall, which characterize  the ordered phase,  are obtained. 
The latter value is larger than $d-1$, indicating that the domain wall
is really rough in the $4d$ Ising spin glass phase. 
\end{abstract}
]

\section{Introduction}
Numerical simulations, in particular Monte Carlo (MC)  methods, have 
played a quite 
important role in spin glass (SG) studies\cite{spinglass}. For
example, very large-scale MC simulations have strongly suggested 
the existence of a SG phase transition in three-dimensional Ising SG
systems.\cite{Ogielski,BhattYoung,KawashimaYoung} 
In these studies, 
a cumulant of SG overlap function $q$, so called the Binder
parameter, has frequently used in order to extract critical
temperature $T_{\rm   c}$.
However, the Binder parameter in $3d$ Edwards-Anderson (EA) Ising
models\cite{KawashimaYoung,Marinari98} merely depends on system sizes
below $T_{\rm c}$ as compared to that above $T_{\rm
  c}$\cite{BergJanke}. 
Moreover, unusual size dependence of the Binder parameter is observed
in short-ranged Ising EA model under the magnetic field\cite{Picco97}. 
Consequently, the existence of the SG phase transition under the field 
has still remained unclear. 
In order to settle the issue and make a progress toward
well-understanding of SG picture, we consider other
numerical analyzes to be quite necessary. 

In this direction, some promising ways have recently been proposed
based on non-equilibrium dynamics\cite{MPRR,OzekiIto} and the idea of
non-self-averaging\cite{self-averaging}, but here we pay attention to 
domain-wall renormalization group method (DWRG) originally proposed by 
McMillan\cite{McMillan}.  
The DWRG estimates a singular part of free energy by calculating
the domain-wall free energy, $\Delta F$, which is defined as the 
free-energy difference between periodic- and anti-periodic boundary
conditions (BC). 
In the scaling regime at low temperatures, $\Delta F$ follows a power law
as a function of  
the system size $L$, $\Delta F\sim L^\theta$,  where the 
stiffness exponent $\theta$ is related to the rigidity of the system. 
If the exponent $\theta$ takes a positive value at a temperature, then
the system  stays in an ordered phase. 
On the other hand, a negative exponent means a disordered phase. 
In this sense, the sign of the exponent $\theta$ is an indicator of the 
existence of long range ordering. 
This exponent $\theta$ also characterizes a low-energy excitation in the SG
phase and is predicted to be smaller than
$(d-1)/2$, $d$ being dimensionality,  in the droplet scaling
theory\cite{FisherHuse1,FisherHuse2}. 

The DWRG approach relies on an accurate way for estimating the
free-energy difference between two BCs. 
It is a difficult task in general for MC method to estimate free
energy or entropy. 
Except for the numerical transfer matrix method for Ising models, 
it is therefore usual to integrate over the free-energy derivative,
measured by MC simulations, along a parameter path between a reference 
system and the one of interest. 
As for zero-temperature calculations, various optimization techniques
have been demonstrated to be useful for Ising\cite{Rieger,Hartmann} and vector
spin systems\cite{Maucourt1}. 
These facts restrict so far to rather small sizes  and/or 
at zero temperature. 
In this work we have developed a boundary-flip MC method proposed by
Hasenbusch\cite{MH93} which allows us to estimate 
free-energy difference at a {\it finite temperature} directly from MC
simulation. 
In applying a naive boundary-flip MC method to large systems
and/or at low temperatures, one may encounter a hardly relaxing
problem even in simple models without many meta-stable states, namely
the system is trapped into a local area in the phase space. 
The original work\cite{MH93} has successfully overcome the
relaxational problem by combining the method with the cluster MC
dynamics. 

In the present paper, we have proposed an alternative strategy, which
is the boundary-flip method with exchange MC (EMC)  method\cite{HN},
in order to 
make the relaxation faster. 
This combined method is found to be quit efficient for
randomly frustrated spin systems such as spin glasses, while the
original method 
based on the cluster MC method is restricted to non-frustrated
systems. 
The present  method is applicable to a wide class of spin systems. 
Moreover, the direct measurements have an advantage over the
thermodynamic integration method from a numerical standpoint, because 
statistical error is controlled within MC scheme in the former. 
Consequently, we have succeeded to estimate the free-energy
difference in a SG model, accurately  enough to observe 
systematic correction to finite-size scaling. 

For applying DWRG to SG systems, we need to choose the relevant
boundary conditions to the ordered phase. The standard approach has
often used a randomly fixed  spin boundary condition\cite{McMillan2}. 
Instead, we employ the replica boundary condition proposed by
Ozeki\cite{Ozeki}, in which two real replicas are coupled with each
other through a boundary surface. 
The replica boundary condition provides that the domain-wall free
energy becomes positive at any bond disorder, implying that it is
conjugates to the SG ordering. This positivity is of benefit to us for 
estimating the domain-wall free energy accurately from a numerical
point of view.  

Here we study the $4d \pm J$ Ising SG model under the replica BC by
the novel MC method. We obtain the critical temperature and the exponent 
by finite-size-scaling analysis of the domain-wall free energy, in
agreement with the previous works. In addition, we estimate two
exponents, the stiffness exponent $\theta$ and the fractal dimension
$d_s$ of the domain wall.  We find that $d_s$ is larger than $d-1$ in
the SG phase.

This paper is organized as follows: 
In the next section, we explain the method for calculating the
domain-wall free energy. 
Section \ref{sec:condition} is mainly devoted to  discussion about 
the replica boundary conditions. 
We give an interpretation of domain wall appearing in the replica
boundary condition and propose a way to measure the morphology of the
domain wall. 
We show results for application of the method to $4d \pm J$ Ising SG
model in the section \ref{sec:result}. 
In the last section, possible extensions of the method and nature of
the low-temperature phase are discussed. 
Appendix A contains a way for setting temperature points which is
needed before simulation in the exchange MC method.

\section{Boundary flip MC method with exchange process}
\label{sec:method}

\begin{figure}[t]
\label{fig:meta}
\epsfxsize=\columnwidth\epsfbox{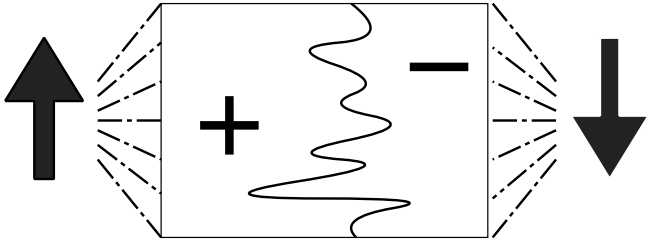}
\caption{
Typical example for metastable configuration in a 
  ferromagnetic model.}
\end{figure}

In this section we describe a method that allows us to evaluate
directly the domain-wall free-energy using MC simulations.
For the sake of simplicity, we restrict ourselves to Ising spin systems
and fixed spin boundary conditions. 
Let us consider a total model Hamiltonian  defined by  
\begin{equation}
  \label{eqn:model}
  {\cal H}_{\rm tot}(\sigma,S_1,S_2) = {\cal H}_{\rm model}(\sigma) + 
  \alpha{\cal H}_{\rm BC}(\sigma,S_1,S_2), 
\end{equation}
where $\sigma$ denotes Ising  spin variable  defined on a $d$-dimensional
hyper-cubic lattice $V$ of $L^d$ and
two additional spins $S_1$ and $S_2$ represent boundary spins. 
The second term gives a coupling between the model system and
the boundary spins along one direction as 
\begin{equation}
  \label{eqn:bc-model}
  {\cal H}_{\rm BC}(\sigma,S_1,S_2) =
  -\sum_{i\in\partial_1 V}J_{i,1}\sigma_iS_1
  -\sum_{i\in\partial_2 V}J_{i,2}\sigma_iS_2,
\end{equation}
where the summation runs over one surface $\partial_1 V$ of the
lattice $V$
and its opposite surface $\partial_2 V$. 
A standard periodic boundary condition  for $\sigma$ is used
along the remaining directions. 
Then the total partition function $Z_{\rm tot}$ and the free energy
$F_{\rm tot}$ of this whole system
are defined by 
\begin{eqnarray}
  \label{partition}
  Z_{\rm tot}(T) &  =  & {\rm
    Tr}_{\{\sigma ,S_1,S_2\}}
\exp (-H_{\rm tot}(\sigma, S_1, S_2)/T) \nonumber\\
  & = & \exp (-F_{\rm tot}(T)/T),
\end{eqnarray}
where $T$ is temperature and we set the Boltzmann constant  to unity.
The phase space of the total Hamiltonian is enlarged by adding the
degree of freedom of the boundary spins  $S_1$ and $S_2$. 
When these  spins are in parallel, the boundary condition is regarded as {\it
  periodic} and similarly the {\it anti-periodic} boundary condition
corresponds  to anti-parallel boundary spins.  
For a given temperature the probability for finding the
periodic boundary condition is given by 
\begin{eqnarray}
  \label{eqn:prob-a}
  P_{\rm P}(T) & \equiv &  \frac
{  {\rm Tr}_{\{\sigma,S_1,S_2\}}\delta_{S_1,S_2}\exp (-{\cal H}_{\rm
    tot}(\sigma,S_1,S_2)/T)}{Z_{\rm tot}(T)}, \nonumber \\   
 & = & \frac{Z_{\rm P}(T)}{Z_{\rm tot}(T)},
\end{eqnarray}
where $\delta$ is the Kroneker delta function. 
This quantity is accessible from MC simulation, namely it is nothing but
the probability for realizing the periodic BC during 
MC simulation in which the boundary spins as well as the bulk spins
are updated according to a standard MC procedure. 
In terms of the probability and the corresponding one to the
antiperiodic BC,  domain-wall free-energy $\Delta F$  we want to
investigate is given by 
\begin{equation}
  \label{free-e}
  \exp(\beta\Delta F(T)) = e^{-\beta(F_{\rm P}-F_{\rm AP})} = 
  \frac{Z_{\rm P}}{Z_{\rm AP}}=\frac{P_{\rm P}(T)}{P_{\rm AP}(T)}.
\end{equation}
This is the basic idea of the boundary flip MC method proposed by
Hasenbusch\cite{MH93}. 
When we adopt a naive local updating process for the boundary spins in 
the boundary-flip MC method, however, we are at once faced to a hardly
relaxing  
problem. For example, once the anti-periodic boundary conditions and
the  domain-wall structure in the system are realized  in the
simulation at low temperatures, as shown in Fig.~\ref{fig:meta},  
the boundary spins are kept to be fixed in the sense that 
the probability for flipping these spins is vanishing in practice. 
This fact makes  statistical error of $\Delta F$ significantly
large. The 
original work\cite{MH93} has overcome this difficulty by utilizing the 
modified cluster flip. 
We can also practically solve this so-called hardly relaxing problem using
recently proposed extended ensemble methods such as the
multicanonical MC method\cite{Berg},  the simulated
tempering\cite{MarinariParisi} and the exchange MC
method\cite{HN}. 
In fact, a similar difficulty has been overcome using the
multicanonical idea in the lattice-switch MC method\cite{Bruce},
which has been proposed to estimate the free-energy difference between
two different crystalline structure in a hard sphere system.  

In the present work, we employ the exchange MC method (EMC) in order
to obtain an efficient path between two boundary-condition states.  
In  the EMC method, we simulate a combined system which consists of
non-interacting $M$ replicated system. The $m$-th replica is simulated 
independently with its own external variable such as temperature. 
We introduce exchange process between configurations of two of $M$
replicas with the whole combined system remaining in equilibrium. 
One possible way for obtaining the path is that we distribute various
values of the coupling $\alpha$ in eq. (\ref{eqn:model}) ranging 
from 0 to 1 to $M$ replicas.   A target system we
are physically interested in is the replica with $\alpha=1$. 
For a replica with null coupling of $\alpha$, which we call a 
source system, the boundary spins can
be flipped freely. Therefore, the path between different 
boundary condition states in the target system would 
be recovered by the exchange process through the source system. 
\begin{figure}
\label{fig:linea}
\epsfxsize=\columnwidth\epsfbox{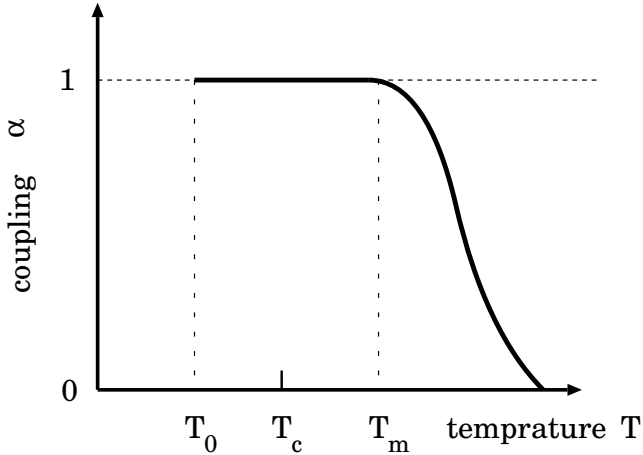}
\caption{
Schematic picture of the exchange line in parameter space. 
}
\end{figure}

In randomly frustrated spin systems such as SG models, there is
another serious relaxation problem arising from bulk spins in the
model system itself. This problem can be overcome also by the EMC
method.\cite{HN} 
When we distribute $M$ temperature points widely including high
enough temperature  in a disordered phase, 
configurations at low temperatures are expected to be refreshed
through the exchange process.  
The EMC method has turned out to work efficiently in the SG
systems\cite{HN,Marinari}. 
Therefore, for the boundary-flip MC method on SG models, 
 we need to construct the EMC method  in two-dimensional
parameter space of the coupling $\alpha$ and the temperature $T$. 
It is possible to introduce the exchange process 
in the two parameter space,  but it is quite time consuming. 
In the present work, therefore, 
we choose an exchange line in the two dimensional space
appropriately, namely we set a system at high temperature with
$\alpha=0$ as one end of the exchange line and systems at lower
temperature with $\alpha$ being unity, as shown in
Fig.~\ref{fig:linea}.  
It is noted that the parameter region of the our final interest lies
on the line with $\alpha=1$ around $T_{\rm c}$ and below. 
An efficient choice of the exchange line would
depend on systems we want to investigate. 
Actual implementation  to the Ising spin glass model will be
explained in detail in Section \ref{sec:result}. 

\section{Replica Boundary Conditions for SG systems}
\label{sec:condition}
In this section, we discuss how to choose boundary condition for SG
systems in the DWRG study.
We concentrate ourselves on a way of choice of boundary condition
along one direction while the remaining ones are considered to be given
appropriately. 
In a conventional DWRG studies\cite{McMillan,BrayMoore}
 as well as the defect energy method, 
a  boundary condition frequently used
is a connected spin BC in which the corresponding boundary term 
in eq. (\ref{eqn:model}) is described by 
\begin{equation}
{\cal H}_{\rm BC}=\sum_{i\in\partial_1 V,j\in\partial_2 V}
J_{ij}\sigma_i\sigma_j. 
\label{eqn:cBC}
\end{equation}
The case with $\alpha/|\alpha |=1(-1)$ is regarded as the 
(anti-) periodic boundary condition. 
For the boundary condition defined by eq. (\ref{eqn:cBC}), 
 the boundary-flip MC method can be applied by 
treating the sign of the coupling $\alpha$ as a MC dynamical variable. 
In SG systems, the free-energy difference between  these BCs cannot be
assured positive so that the width of distribution of the 
free-energy difference is examined as an  
effective coupling of the SG ordering, $F_{\rm eff}=\sqrt{(F_{\rm
    AP}-F_{\rm     P})^2}$. 
To evaluate the mean width is rather difficult as compared with the
average in numerical calculations. 
Further, it is less clear how the domain wall is created in a 
random spin system under these BCs. 

In order to avoid the difficulty and make clear an idea of the domain 
wall, Ozeki\cite{Ozeki} has proposed a replica boundary condition
(RBC), in which two real replicas are prepared with the
same bond realization.  
Its essential point is to introduce a uniform coupling between these
two replicas  only for one surface $\partial_0 V$ along a given direction. 
For the other directions  periodic BC is employed  as usual. We show
explicitly an example expressed as the Ising Hamiltonian   
\begin{equation}
  \label{eqn:pmJmodel}
  {\cal H}_{\rm model}(\sigma,\tau) = -\sum_{\langle
    ij\rangle}J_{ij}(\sigma_i\sigma_j+\tau_i\tau_j)
 -J_{\rm int}\sum_{i\in\partial_0 V}\sigma_i\tau_i, 
\end{equation}
where both $\sigma$ and $\tau$ are Ising variables and the summation 
of the first term runs over nearest neighbor bonds. 
The second term corresponds to the replica interaction mentioned above. 
When $J_{\rm int}$ is set to (anti-) ferromagnetic, the boundary
condition is called replica (anti-) periodic, R(A)PBC. 
Spins on the opposite side of  $\partial_0 V$  are kept randomly
fixed with $\sigma_i=\tau_i$. 

Ferromagnetic interactions between the replicas in the RPBC prefer
a {\it   self-overlap state}, even if the system has many local minima
or pure states.  Namely, one replica gives an effective conjugated field
to the other replica through the inter-replica interaction.
It is convenient to consider the domain wall in terms of the replica
overlap $q_i=\sigma_i\tau_i$. 
The self-overlap state is characterized by positive values of
$q_i$ at all the sites, meaning no domain wall in the system. 
At sites  on the opposite surfaces of $\partial_0 V$, $q_i$ take unity
by definition, irrespective of R(A)PBCs. On 
the other hand,  
anti-ferromagnetic inter-couplings between the replicas in the RAPBC
would induce negative overlap at sites near the coupling. Therefore,
at least one domain wall, characterized by a region where the sign of
$q_i$ changes, likely appears in the RAPBC, if the system 
has a rigid ordered state. 
From a mathematical point of view, 
non-negativity of the free-energy difference $\Delta F_{\rm
  R}=F_{\rm RAPBC}-F_{\rm RPBC}$ under the replica BC has been  proven 
rigorously in any random Ising model at any finite temperature using
the transfer matrix formalism\cite{Ozeki}. 
This  non-negativity holds true 
irrespectively of a choice of spins on the surface opposite to
$\partial_0 V$.
As a result, only average of the domain-wall free energy is needed for
estimating an relevant effective coupling of 
the SG ordering. 
This is advantageous for reducing statistical error of $\Delta
F_{\rm R}$ from which  a transition point from 
paramagnetic to SG phase is detected. 

An additional merit of the replica boundary condition is that  
we can discuss the morphology of the domain wall at finite
temperatures. 
In terms of the local overlap $q_i$, the area of the domain boundary
mentioned above is expressed as $W=\sum_{\langle
  ij\rangle}\frac{1}{2}(1-q_iq_j)$, where the summation  is over
nearest-neighboring pairs. . 
Then we can extract directly domain-wall properties such as its
fractal dimension, from the difference $\Delta W(T)$ defined by 
\begin{equation}
  \Delta W(T)=\frac{1}{2}\sum_{\langle ij\rangle}
(\langle \sigma_i\sigma_j\tau_i\tau_j\rangle_{\rm RPBC}-
\langle \sigma_i\sigma_j\tau_i\tau_j\rangle_{\rm RAPBC}), 
\label{eqn:domain-wall}
\end{equation}
where $\langle\cdots\rangle_{\rm R(A)PBC}$ denotes the thermal average 
under the replica (anti-) periodic BC. 
This quantity is also regarded as a difference of link
correlation\cite{Ciria} between two boundary conditions in $\pm J$ models.  
The correlation function as well as the replica overlap have been
studied in a similar replicated system\cite{Ciria} which has a
global coupling between the replicas. This coupled system is different
from the present system  under RBC. 
In particular, the correlation function (\ref{eqn:domain-wall}) is
related to domain-wall properties only in the RBC.  
The domain-wall area $\Delta W$ has not been directly studied so far in SG
systems, except for the zero temperature calculation in two
dimensional Ising SG model\cite{BrayMoore2}.  
We will present new results for $\Delta W$ in the next section.

\section{results}
\label{sec:result}
In this section, 
we present results of an application of the novel MC method  explained 
in the previous sections to $4d \pm J$ Ising SG model.  
The interactions $\{J_{ij}\}$ in eq. (\ref{eqn:pmJmodel}) are random
variables which take values $\pm J$ with equal probability.
The boundary-flip MC method can be applied to the replica BC by
regarding the sign of the interaction $J_{\rm int}$ in
eq. (\ref{eqn:pmJmodel})  as a dynamical variable. 
Equivalently these boundary conditions are defined 
by relative direction of the boundary spins $S_1$ and $S_2$  
added to eq. (\ref{eqn:pmJmodel}) whose $J_{\rm   int}$ are fixed to
be positive.    
Then, the boundary part in eq. (\ref{eqn:model}) is given by 
\begin{equation}
  \label{eqn:model4d-bc}
  {\cal H}_{\rm BC}(\sigma,\tau,S_1,S_2) = 
  -\sum_{i\in\partial V}J_i(\sigma_iS_1+\tau_iS_2) ,
\end{equation}
where the interactions $J_i$ are also distributed randomly.  
In the present work we adopt this method with the boundary spins. 

The number of the replicas  $M$  in the EMC method  is fixed $32$
irrespectively of the system sizes to utilize the multi-spin coding
technique. Each replica with the parameters $\alpha$ and $T$ tries to
exchange configuration with the nearest replica in the parameter
space.  
As we have explained in Section \ref{sec:method}, we choose in this
two-parameter space, a line on which $M$ replicas are prepared. The
line chosen is such that  
value of $\alpha$ is unity below a certain temperature $T_m$, but
it decreases like a Gaussian formula as a function of $T-T_m$ above $T_m$. 
The onset $T_m$ is set to be about 2 times of the critical
temperature. 
We distribute the set of the parameters to the 32 replicas such that
the acceptance ratio for each exchange process becomes independent of
the replicas. This can be succeeded by a simple iteration
method using the energy function, which is estimated from
a short preliminary  run.
Details of the iteration method is explained in Appendix. 

\begin{figure}
\label{fig:eq}
\epsfxsize=\columnwidth\epsfbox{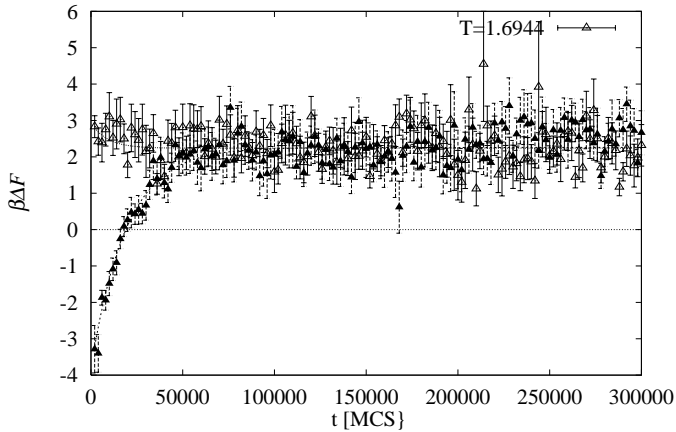}
\caption{
The domain-wall free-energy of $4d \pm J$ Ising SG model  
with $L=8$ and $T=1.694$ well below the SG transition temperature 
as a function of MC steps. 
The upper data marked by open triangle are started from the periodic
boundary condition for the whole system, while the lower one from the
anti-periodic one. 
Each point at the time $t$ is obtained by averaging over 2,000 MCS
around $t$ and error bars  are  estimated from statistical fluctuation 
 over 10 samples. }
\end{figure}

As an  equilibration check, we study time evolution of $\Delta F_{\rm
  R}$ starting from two initial conditions: periodic and anti-periodic
boundary conditions 
imposed for the whole replicated systems in the EMC simulation. 
The initial conditions for the bulk spins are chosen at random. 
The free-energy difference $\Delta F_{\rm R}$ is estimated as a
function of MC  step $t$ by averaging over short MC steps around the
time $t$.  In the case of the whole anti-periodic BC, free-energy
difference,  starting from a large negative value at the initial time,
evolves toward equilibrium. 
The other estimation with the periodic BC at the initial time 
reaches  to the equilibrium value from the opposite direction to that
of the anti-periodic BC. 
In equilibrium, two curves coincide with each other.  
As expected, we see in Fig.~\ref{fig:eq} that the equilibration
of $\Delta F_{\rm R}$ is obtained after a certain time.
It should be noted that the relaxational function approaching to the
equilibrium value follows an exponential law rather than a power law
observed in the standard SG simulations. This implies the existence of
a typical 
time scale for equilibration in the present  method. 
We thus expect that the system really reaches equilibrium after a
few times of such time scale. 
We estimated the time scale for  other sizes  and determined the 
MC steps (MCS)  for thermalization and measurements. For example, 
in simulations of $4d$ case  with $L=8$, we take $9.6\times 10^4$
MCS for the initial step  and $2.0\times 10^5$  MCS for measurement. 
We have also checked that the ergodic time\cite{Berg,HN} is 
 about $3\times10^2$, $3.0\times10^4$, $5.8\times10^4$ and
 $1.7\times10^5$ MCS on average for $L=4, 6, 8,$ and $10$,
 respectively. 

\begin{figure}
\label{fig:delfh00}
\epsfxsize=\columnwidth\epsfbox{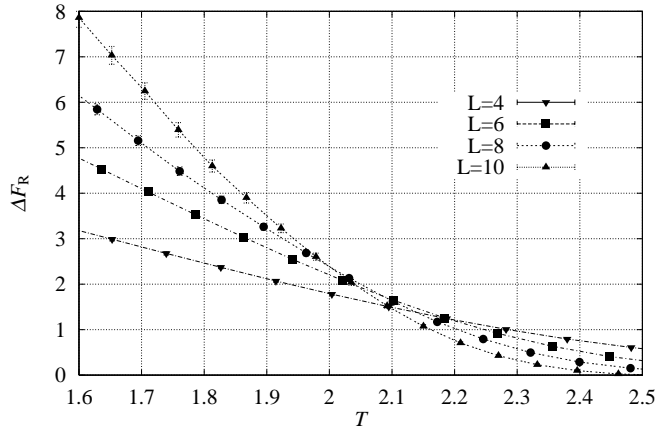}
\caption{
Temperature dependence of the domain-wall free-energy for $4d$
  $\pm J$ Ising SG model near the critical
  temperature. These lines are for guide of eyes. 
}
\end{figure}

\begin{figure}
\label{fig:scal-df}
\epsfxsize=\columnwidth\epsfbox{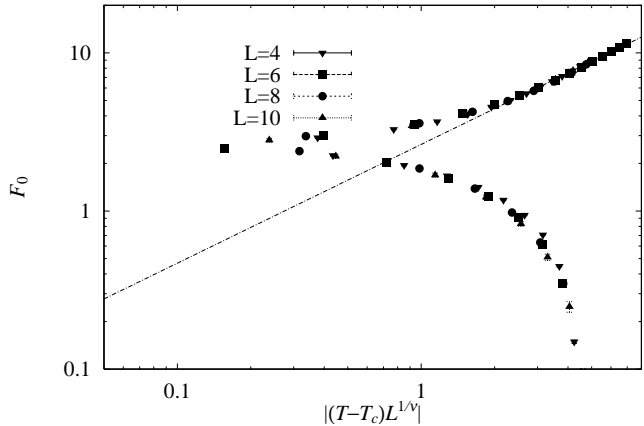}      
\caption{
Finite-size scaling plot of the domain-wall free energy in $4d \pm J$
Ising SG  model. The leading correction to the scaling is taken into
account.  The scaling plot after subtraction of the leading correction
is shown.  The estimated scaling parameters are $T_c=2.00(4)$,
$\nu=0.92(6)$ and the  irrelevant exponent $\omega=1.5(9)$. The
slope of the scaling function is  
asymptotically close to $0.75(1)$, meaning that the stiffness exponent
$\theta$ is  $0.82(6)$.
}
\end{figure}

We show  temperature dependence of  $\Delta F_{\rm R}$ for the $4d$
Ising SG model in Fig.~\ref{fig:delfh00}.  
The lattice size studied are $L=4, 6, 8,$ and $10$ with samples 2197,
 2060, 1332, and 892,  respectively. 
According to the standard finite-size-scaling argument, the
domain-wall free energy should be scaled as 
\begin{equation}
  \label{eqn:scal-delf}
  \Delta F_{\rm R}(L,T) \sim F_0((T-T_{\rm c})L^{1/\nu}), 
\end{equation}
where the parameter $\nu$ denotes  the critical exponent of the
correlation length and $F_0$ is a scaling function. 
Therefore, the critical temperature can be located in the point where
$\Delta F_{\rm R}$ for different sizes as a function of $T$ cross
with each others. 
The crossing feature of $\Delta F_{\rm R}$ at $T_{\rm c}$ is common to 
the Binder parameter. 
In fact, as shown in Fig.~\ref{fig:delfh00}, crossing of $\Delta
F_{\rm R}$ of two different sizes is seen 
at a certain  temperature. However the crossing temperature is
found to shift systematically to low temperature side as the system
size increases, implying that correction to the finite-size scaling
is significant.  
We consider correction due to the leading irrelevant scaling
variable whose scaling dimension is $\omega$;  
\begin{equation}
  \label{eqn:correction}
  \Delta F_{\rm R}(L,T) \sim F_0((T-T_{\rm c})L^{1/\nu})+ L^{-\omega}
  F_1((T-T_{\rm c})L^{1/\nu}),   
\end{equation}
These exponent $\nu$ 
and $\omega$ and the critical temperature $T_c$ are determined by
fitting the simulated data to the scaling formula
(\ref{eqn:correction}), where the scaling functions $F_0$ and $F_1$
are assumed to be given by third order polynomial functions. 
From the fitting, we estimate $T_{\rm c}=2.00(4)$, $\nu=0.92(6)$, and 
$\omega=1.5(9)$. 
The finite size scaling of $F_0$ after subtraction of the leading
correction is plotted 
in Fig \ref{fig:scal-df}, where all the data points are found to
collapse almost into a universal function. 
The scaling plot including the smallest size $L=4$ is obtained only
when the leading term of the correction is taken into account. 
The estimated critical temperature is consistent with 
the previous results obtained  by the MC method\cite{Badoni}
 and the high temperature expansion\cite{Singh,KAAHM}. 
Our result for $\nu$ is also in agreement with these expansion
studies, and not very different with that obtained by MC simulations
for $\pm J$
\cite{Badoni} and Gaussian distribution
\cite{PTR}.  Since 
the system sizes used in the present work are larger than those in the
previous MC simulations, we expect that our estimation is reliable. 
The irrelevant exponent $\omega$ is, to our knowledge, the first
estimation for $4d$ Ising SG model by MC simulation,  
but its value is slightly lower than that obtained from the series
expansion\cite{KAAHM} that quoted about 3. 

\begin{figure}
\epsfxsize=\columnwidth\epsfbox{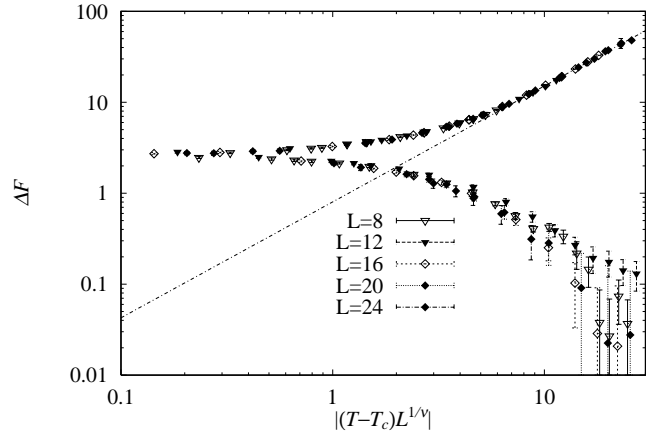}
\caption{
Finite size scaling plot of the domain-wall free energy in $3d$ ferromagnetic 
Ising model. The parameters of the scaling are estimated as follows: 
$T_c=4.5117(4)$, $\nu=0.624(7)$. 
The asymptotic behavior of the scaling function follows a power law as 
a function of the scaling parameter $(T-T_c)L^{1/\nu}$ with slope $1.27(1)$. 
The value of the slope is compatible with the low temperature behavior, namely
$\theta\nu$, being  $\theta=d-1$. 
\protect{\label{fig:df3d}}
}
\end{figure}

At low enough temperature, the domain-wall free energy is expected to
be scaled as 
\begin{equation}
  \Delta F_{\rm R}(L,T) \sim L^{\theta},
\end{equation}
where $\theta$ is an exponent which gives the characteristic energy
scale $L^\theta$ of low energy excitations of typical size $L$.
We cannot evaluate $\Delta F_{\rm R}$ at low temperatures enough to
distinguish the low temperature properties from the critical
behavior. 
Here we try to estimate the exponent $\theta$ from the scaling
function of $\Delta F_{\rm R}$. 
We assume that the behavior of $\Delta F_{\rm R}$ at a large length
scale is also described by the scaling form of eq. (\ref{eqn:correction})
near below $T_{\rm c}$.  This assumption implies that 
the asymptotic behavior of
the scaling function $F_0$ is predicted as 
\begin{equation}
  \label{eqn:asymptotic}
  F_0(x) \sim |x|^{\theta\nu}, 
\end{equation}
at $x\rightarrow -\infty$. 
We examine this scaling idea in the simple $3d$ Ising ferromagnetic
model, where the stiffness exponent coincides with the surface
dimensions $d-1$. 
We estimate the domain-wall free energy by the present MC method
under the connected spin BC described in (\ref{eqn:cBC}). 
In the $3d$ Ising model, we scale the data to the leading scaling
formula (\ref{eqn:scal-delf}) without the correction, because we have 
not observed a shift of the crossing temperature under our numerical
accuracy. 
The finite-size scaling of the domain-wall free
energy works well as observed in Fig.~{\protect{\ref{fig:df3d}}}. 
The asymptotic behavior of the scaling function gives
$\theta\nu\sim1.27$, compatible with the well-known values of $\nu$
and $\theta=d-1$. 

Let us turn to the $4d$ Ising SG model. 
The stiffness exponent $\theta$ in SG systems is expected much smaller 
than that of the ferromagnetic model.  The droplet theory predicted
the upper bound of $\theta$ to be $(d-1)/2$\cite{FisherHuse2}. 
We extract value of $\theta$ from the scaling function obtained in
Fig.~\ref{fig:scal-df}. 
We fit  the scaled data 
with the scaling variable $x$ larger than 3 to a power law. 
The best fit is obtained with the exponent $\theta\nu=0.75(1)$, which
yields the stiffness exponent of $\theta=0.82(6)$. 

\begin{figure}
\label{fig:dw4d}
\epsfxsize=\columnwidth\epsfbox{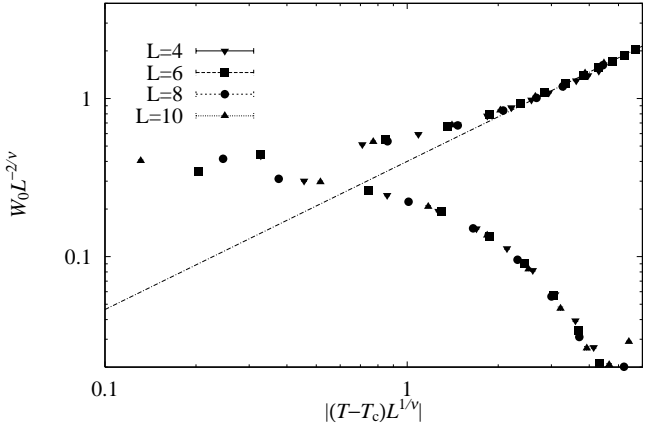}
\caption{
Finite-size scaling plot of the domain-wall area in $4d \pm J$ 
Ising SG model after subtraction of the leading correction to the scaling. 
The critical temperature is used the result of the scaling analysis for the
 domain-wall free energy. The exponent $\nu$ is found to be $0.94(2)$, 
consistent with the previous estimation. The estimated irrelevant exponent 
$\omega=1.86(77)$ agrees with that obtained from the $\Delta F_{\rm R}$
scaling. The slope is estimated $0.94(2)$, suggesting 
$d_s=3.13(2)$. 
}
\end{figure}

We also investigate the domain-wall  area $\Delta W$ defined by
eq. (\ref{eqn:domain-wall}) in this model, which is easily calculated in
the present MC scheme. 
A scaling analysis similar to the one for $\Delta F_{\rm R}$ is performed for
$\Delta W$, taking into account the leading correction to the
scaling. It is noted that in contrast with the $\Delta F_{\rm R}$
scaling, $\Delta W$ is proportional to $L^{2/\nu}$ near $T_{\rm c}$
because it 
has  essentially the same scaling dimension as  the energy-energy
correlation function.  
The finite-size-scaling plot for $\Delta W$ is shown in
Fig.~{\ref{fig:dw4d}}, where the critical temperature is used 
which is estimated  by the $\Delta F_{\rm R}$ scaling. 
The scaling nicely works both above and below $T_{\rm c}$ and the
estimated $\nu$ value is consistent with that from $\Delta F_{\rm
  R}$. 
We suppose that at low temperature the domain wall 
in the SG system is rather rough. Correspondingly 
the domain-wall area $\Delta W$ is  expected to follow a power law 
on size with  a non trivial fractal dimension $d_s$. 
We estimate $d_s$ by extracting the asymptotic behavior of the
scaling function of $\Delta W$ in the same way as in the analysis of
$\Delta F_{\rm R}$. 
The asymptotic slope of the scaling function is $d_s\nu-2$. 
The fractal dimension of this model is found to be
3.13(2). 
According to the Bray--Moore scaling law\cite{BrayMoore2}, the
exponents $\theta$ and $d_s$ are related
to the chaos 
exponent $\zeta$   
\begin{equation}
  \zeta = \frac{d_s}{2}-\theta. 
\end{equation}
By this combined with the values of $\theta$ and $d_s$ obtained here,
our estimation of $\zeta$ is 0.75(6). 
This value is smaller than those of MC simulations for $4d$ Ising SG
models\cite{NeyNifle,Azcoiti}, but rather close to that by
the Migdal-Kadanoff renormalization group analysis\cite{NifleHilhorst}. 

\section{Discussion and Summary}
\label{sec:discussion}
We have developed a numerical method which enable us to estimate
free-energy difference directly from MC simulation. It is a boundary
flip MC method, in which the replica boundary conditions and the exchange
MC technique are incorporated. 
 The proposed method
works well in the short-range Ising SG model. 
This method presented here can be applied to various spin systems 
including vector spin models  
because our argument does not depend on model Hamiltonian. 
It should be noted that the EMC method, as well as other extended
ensemble methods, is also applicable to randomly
frustrated spin systems, while the cluster-flip based method is
restricted in non-frustrated models. 
Another extension would be concerned with the choice of the boundary
conditions. 
In this paper, we have described the case for the fixed spin BC, but 
it is straightforward to extend it to other type of BCs. 
It is only necessary for boundary conditions to be expressed by a
countable variable,  while the degree of freedom
of the model system is not restricted. 

We also discuss boundary condition for SG systems. 
Let us comment on related studies. 
A similar coupled replica system has been studied analytically
by mean-field variational method\cite{Franz}, where two replicas are
coupled with each other by fixing the value of overlap between surface 
spins of these replicas. 
The system studied roughly corresponds to the present replica
boundary model by choosing appropriate parameters. 
It is predicted that an excess free energy due to the effective
coupling is  proportional to $L^{d-5/2}$, which accidentally
coincides with the upper limit of the droplet scaling theory in the
four dimensional case.  
Our estimation of the stiffness exponent is not compatible to that predicted
from the variational calculation. 

Recently a new boundary condition, called the naive boundary condition, 
 has been proposed in 2D
Ising\cite{Shirakura} and XY\cite{Kosterlitz} 
spin glass models, independently.
In these studies, they minimize energy of a whole system under the
free boundary condition.  
Using the  obtained boundary spin configuration  as a reference
system,  
a twisted boundary condition is prepared by flipping the sign of spins on
one surface. The ground state energy of such system is always higher than
that of the reference system. 
They claimed that this non-negativity is an evidence of introducing
correctly a domain wall into the system. 
It is doubtful whether such boundary conditions defined at zero
temperature is also relevant to the ordering at finite temperatures. 
This is because many SG systems including both short range\cite{BrayMoore2}
and mean field models\cite{Kondor} 
are expected to exhibit chaotic nature; namely spin configurations at
finite temperatures  differ from those at $T=0$ in larger scale than
the so-called overlap length. 
Further, the replica boundary condition takes an advantage from
the naive one in a practical sense, because the former does not need the
ground state calculations. This fact makes our investigations easier
in three or high dimensional systems, where the ground states are
hardly found for suitable large system due to NP hardness.

The present method has successfully  been applied to the $4d \pm J$
Ising  SG model under the replica boundary conditions. The average of
the domain-wall free energy $\Delta F_{\rm R}$ over samples, not the
variance as used in the standard DWRG study, exhibits very clear
crossing at the critical  temperature, implying that it is a good
indicator of the SG transition. It is noted that the replica BC is
crucial for providing the non negativity of $\Delta F_{\rm R}$. 
We  expect that,  when the system has a well defined rigidity in the
ordered phase, the $\Delta F_{\rm R}$ analysis works well even in the
case  where the Binder parameter does not show a crossing at $T_{\rm c}$. 
In such systems, the short range SG models with the field are one of
the most attractive problems in the SG study.  
As a byproduct of the RBC, we can argue the domain-wall
area in the SG phase.  We have estimated the stiffness exponent
$\theta$ and the surface dimension $d_s$ of the domain wall in the
$4d$ Ising SG phase independently. The latter value lies significantly 
above the trivial surface dimension $d-1$, meaning that the domain wall is
rough, while both $\theta$ and $d_s$ coincide with $d-1$ in the
ferromagnetic Ising models. 

\begin{figure}
\epsfxsize=\columnwidth\epsfbox{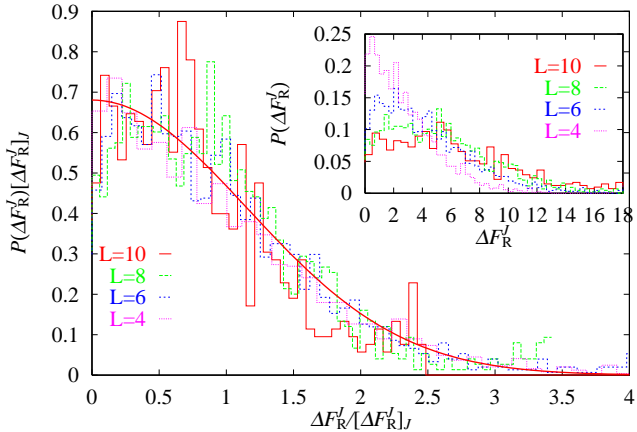}
\caption{Scaling plot of the distribution function of the
  domain-wall free energy with $T\sim 1.6$. Both axes are scaled by
  the first moment of the distribution. The solid curve is obtained by 
  fitting the scaling function to a Gaussian formula. The raw data is
  shown in the inset.
\protect{\label{fig:hist-scal}}
}
\end{figure}

Finally we make a comment on distribution of $\Delta F_{\rm R}$
over samples $P(\Delta F_{\rm R})$, whose typical results are 
shown in Fig.~\ref{fig:hist-scal}. 
To our surprise, 
the distribution functions of different sizes, when scaled by their
first moment, lie top on each others in the SG phase. 
Another remarkable observation is that 
the scaling function is approximated by a Gaussian function;  namely 
it approaches to a nonzero value as its argument goes to zero. 
These results, similar to those observed in $2d$ and $3d$ Ising SG
models at zero temperature\cite{BrayMoore,BrayMoore86}, are
consistent with the droplet picture\cite{FisherHuse1,FisherHuse2}. 

The question of whether many equilibrium pure states exist
or not in the SG phase has still remained controversial. 
For the system of present interest, some MC studies\cite{Reger,Ciria}
have supported the existence of the multiple pure states, namely the
mean field  picture, while the Migdal-Kadanoff approximation for the
short range SG 
model\cite{Moore} has  claimed that the asymptotic size scale to detect
the correct thermodynamic properties is far from those investigated
in the MC simulations.  
As mentioned in Sec.~\ref{sec:condition}, the replica BC used in the
present work prefers a 
self-overlap configuration in the two replicas. 
Correspondingly, under the replica antiperiodic BC, 
there likely appear such configurations with a domain wall which
lies in one of the two replicas and separates one configuration  from
its time-reversal one. 
Therefore, our results mentioned above strongly suggest that 
nature of low-lying excitations within one pure state is as expected
in the droplet theory.  Our data along, however, cannot exclude the
possibility that there are many pure states.

In conclusion, we have proposed a MC method which enable us to
estimate the free-energy difference and 
have  successfully applied it to $4d \pm J$  Ising SG model. 
Our value of $T_{\rm c}$ is in good agreement of the previous results
obtained from the numerical simulations and the series expansions. 
We have presented estimates of two exponents, the stiffness
exponent and the fractal dimension. 
We have also found that low-lying excitations as expected 
in the droplet theory are realized within one pure state in the SG
phase, though we cannot rule out  the possibility that 
there exist many pure states. 

\acknowledgments
The author would like to thank H.~Takayama for valuable
suggestions and a critical reading of the manuscript. 
He also thanks Y.~Ozeki, H.~Yoshino and S.~Todo for helpful
discussion. 
Numerical calculation was mainly performed on DEC alpha personal
workstations and Fujitsu VPP500 at the supercomputer center, Institute
of Solid State Physics, University of Tokyo. 
He used the internet random number server RANSERVE made by S.~Todo to get 
the initial set of random numbers in MC simulations. 

\appendix
\section{Setting temperature points for the exchange MC method}
In this appendix  we propose a practical way 
to determine temperature set which is needed in the exchange MC
method. 
For simplicity, we consider a procedure for setting  a temperature
point $\beta_n$ between  two fixed ones, $\beta_{n-1}$ and $\beta_{n+1}$. 
Our criterion is that acceptance probabilities for the exchange 
trial with both neighboring temperatures become equal: 
\begin{eqnarray}
  (\beta_{n-1}-\beta_n)(E(\beta_{n-1})-E(\beta_n)) & = & C, \nonumber \\
  (\beta_n-\beta_{n+1})(E(\beta_n)-E(\beta_{n+1})) & = & C,
\end{eqnarray}
where $C$ and $\beta_n$ are unknown constants.
A formal solution for $\beta_n$ is given by 
\begin{eqnarray}
&  &  \beta_n = g(\beta_n) = \frac{1}{E(\beta_{n-1})-E(\beta_{n+1})}
\times
( \beta_{n-1}E(\beta_{n-1})
\nonumber \\
& &
\quad\quad -\beta_{n+1}E(\beta_{n+1})-E(\beta_n)(\beta_{n-1}-\beta_{n+1})).
\end{eqnarray}
Regarding $\beta'=g(\beta)$ as a map of $\beta$ to $\beta'$, 
we find an fixed point of period 2 
with $\beta_{n+1} = g(\beta_{n-1})$ and $\beta_{n-1}=g(\beta_{n+1})$.
Therefore, we expect a repulsive fixed point between
$\beta_{n-1}$ and $\beta_{n+1}$. A new mapping to obtain the 
fixed point is given by
\begin{equation}
  \beta_n(t+1) = \frac{1}{2}(\beta_n(t) + g(\beta_n(t))), 
\end{equation}
where $t$ is the iteration step.
This iteration scheme can be extended straightforwardly to the case
for multiple temperature points. 
The whole set of temperature is divided into two groups with even-$n$
and odd-$n$. 
Using the iteration scheme, temperature points of the one group are
updated with the other group fixed, alternatively. 
In actual iterations, the initial temperature points $\{\beta_n\}$
are set in a suitable way, for example,  equidistant $\beta$. 
The energy $E(\beta)$ at the initial set of $\beta$ is roughly
estimated by short MC simulation and the energy at any temperature
between $\beta_1$ and $\beta_M$ is assumed to be obtained from the MC
data, for example,  by interpolation technique. 
The convergence of the iteration is rapidly achieved in many 
systems we have investigated. 

From our experiences so far, efficiency of the EMC method is rather 
insensitive for the choice of temperature points, when it is 
applied to systems, such as spin glasses, with non-diverging 
specific heat at the phase transition. This fact that it is not 
necessary to specify any parameters before main simulation is, 
in fact, one of big advantages of the EMC method against the other 
extended ensemble methods such as the multicanonical MC method and
simulated tempering method. 
Nevertheless we emphasize that a little effort on preparing the 
temperature points by pre-MC runs following the prescription 
described above ensures the acceptance ratio almost 
independent of temperature and so is quite useful.

\end{document}